# Theory and computer simulation of the moiré patterns in single-layer cylindrical particles

VLADIMIR SAVELJEV[1,*] AND IRINA PALCHIKOVA[2,3]

[1] *Department of Physics, Myongji University, 116 Myongji-ro, Cheoin-gu, Yongin, Gyeonggi-do, Korea, 17058*
[2] *Technological Design Institute of Scientific Instrument Engineering (TDISIE), Siberian Branch RAS, 41, Russkaya St., 630058 Novosibirsk, Russia*
[3] *Novosibirsk State University, 2, Pirogova St., 630090 Novosibirsk, Russia*
*\*Corresponding author: saveljev.vv@gmail.com*

**Basing on the theory for arbitrary oriented surfaces, we developed the theory of the moiré effect for cylindrical single-layer objects in the paraxial approximation. With using the dual grids, the moiré effect in the plane gratings is simulated, as well as the near-axis moiré effect in cylinders including the chiral layouts. The results can be applied to the graphene layers, to single-walled nanotubes, and to cylinders in general. © 2016**

http://dx.doi.org/10.1364/AO.99.099999

## 1. INTRODUCTION

An optical interaction between two or more transparent or semi-transparent periodical structures often results in the visual moiré effect [1-4]. The moiré effect was many times observed using other rays, for example, in the electron microscope [5-8], where the electron beams are deflected similarly to the light rays in linear optics. For the moiré effect, the periodical structure of the interacting surfaces is important; therefore we refer them to as gratings.

The moiré effect is well investigated, particularly, when the gratings are parallel to each other [9, 10] or when a plane is combined with a curved surface [11, 12].

Due to similarity between the electron microscope and the optical microscope, the nano- and micro-objects are observed in parallel rays, i.e., effectively from the infinite distance, when the paraxial approximation can be used.

In the current paper we consider the moiré effect in non-planar regular curved objects, i.e., in cylinders. Examples of such surfaces at the macro-level are the air- and liquid filters, bent meshed elements of the architectural structures (grid shells); examples at the nano-level are layered and cylindrical nano-particles including the single-walled chiral nano-particles (SWNT).

In simulation and in experiments we assume that the radius of the visibility circle [1] in the spectral domain is shorter than the distance from the origin of the spectral domain to the closest spectral component of the Fourier spectrum of either grating. Also in this paper, we only consider the layout of the moiré patterns, and especially, their period and angle, whereas the amplitude of the moiré patterns is out of scope. Therefore, the lowest and highest intensities of the simulated moiré patterns supposed to be near zero and one.

The paper is organized as follows, In Sec. 2, we present the theory for the arbitrary oriented plane gratings including equiangular ones and for a meshed (semi-transparent) cylinder; the relationships between the dual tesselations (hexagonal and triangular), are also considered. In Sec. 3, we present the computer simulation of the moiré patterns in cylinders with slanted (chiral) orientation of the mesh. The experiments were made with using the gratings printed on the transparent film. In Sec. 4, we discuss the results and finalize the paper.

## 2. THEORETICAL BACKGROUND

### A. Two arbitrary planes

Consider two gratings lying in arbitrary oriented planes. According to the theory [1, 2] for the gratings perpendicular to the line of sight **u**, the wavevector of the moiré wave **k$_m$** is equal to the difference of the wavevectors of gratings. On the other hand, for gratings arbitrary oriented in space, the visual effect is determined by the tangential components **k**$_t$ of the wavevectors of gratings lying in the surfaces $q_1$ and $q_2$ orthogonal to the line of sight (LOS),

$$\mathbf{k}_m = \mathbf{k}_{t1}^{(proj)} - \mathbf{k}_{t2}^{(proj)}. \quad (1)$$

More generally, the moiré wavevector is a linear combination of the tangential spectral components with integer coefficients, which fall within the visibility circle. Equation (1) represents the combination of the first harmonics of gratings with the coefficients +1 and -1.

To visualize the optical interaction, both gratings should be seen through each other or, effectively, projected onto the same plane. Moreover, the second grating projected onto $q_2$ is additionally projected onto the surface $q_1$ of the first grating.

An observer is at the distance $z = l$, while the gratings are at $z = 0$ and $z = -\Delta l$. The vectors $\mathbf{k}_t$ are directed along the corresponding tangential vectors $\mathbf{t}$. The picture of the observation of the moiré patterns along the LOS is shown in Fig. 1.

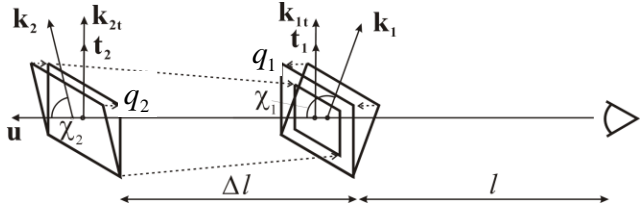

Fig. 1. View along the line of sight.

The projections are shown in Fig. 1 by dotted lines. Note, that projected are physical quantities measured directly, such as the length, distance, size, etc., but not wavevectors. In such projection, the first grating is unchanged, while the period of the projected second grating is divided by the geometric factor according to [13],

$$p_2^{(proj)} = \frac{p_2}{s}, \quad (2)$$

where $s = 1 + \Delta l/l$ is the geometric characteristic of the layout of the tangential planes; in this expression $l$ is the distance from the observer to the first (nearest to the observer) grating and $\Delta l$ is the gap between layers; both $\Delta l$ and $l$ are measured along the LOS, as shown in Fig. 1. The wavevector is an inverse period, and $\mathbf{k}_2$ is re-scaled according to Eq. (2). Then, Eq. (1) is transformed to

$$\mathbf{k}_m = \mathbf{k}_{t1} - s\mathbf{k}_{t2}. \quad (3)$$

For an oblique surface, the modulus of the tangential component $\mathbf{k}_t$ can be found from the geometric relationship between the projected physical quantities, i.e., periods; as follows

$$|\mathbf{k}_t| = \frac{|\mathbf{k}|}{\sin \chi}, \quad (4)$$

where $\chi$ is the angle between the vectors $\mathbf{k}$ and $\mathbf{u}$. The angle $\chi$ defines the dot product of the vector LOS with the wavevector of grating as $\cos \chi = (\mathbf{k} \cdot \mathbf{u})$, thus $\sin \chi = [(1 - (\mathbf{k} \cdot \mathbf{u})^2)]^{1/2}$ and

$$\mathbf{k}_t = \frac{|\mathbf{k}|}{\sqrt{1 - (\mathbf{k} \cdot \mathbf{u})^2}} \mathbf{t}, \quad (5)$$

where $\mathbf{t}$ is the tangential vector of the surface $s$ (perpendicular to $\mathbf{u}$). In Eq. (5), the vectors $\mathbf{k}$ and $\mathbf{u}$ are given, the absolute values of $\mathbf{t}$ and $\mathbf{n}$ are equal to one unit; whereas the direction of $\mathbf{t}$ should be defined. Among infinitely many tangential vectors, we choose the vector lying in the same plane with $\mathbf{k}$ and $\mathbf{u}$. In this case, the vectors $\mathbf{u}$, $\mathbf{k}$, and $\mathbf{t}$ are linearly dependent. It means that $\mathbf{t}$ can be expressed in terms of $\mathbf{k}$ and $\mathbf{u}$.

Once two tangential vectors are found, the moiré vector can be derived from Eqs. (3) and (4),

$$\mathbf{k}_m = \frac{k_1}{\sin \chi_1}\left(\mathbf{t}_1 - \frac{sA}{\rho}\mathbf{t}_2\right), \quad (6)$$

where $\rho$ is the size factor (the ratio of the modules of the wavevectors)

$$\rho = \frac{k_1}{k_2}, \quad (7)$$

and $A$ is the oblique factor

$$A = \frac{\sin \chi_1}{\sin \chi_2}. \quad (8)$$

In the coplanar orthogonal gratings, $s = 1$, $\chi_1 = \chi_2 = 90°$, and Eq. (6) becomes

$$\mathbf{k}_m^c = k_1\left(\mathbf{t}_1 - \frac{1}{\rho}\mathbf{t}_2\right). \quad (9)$$

Particular cases of Eq. (6) can be represented in the scalar form in terms of periods $p$ instead of wavevectors $k$ as follows.

In the plane layout of vectors, three vectors $\mathbf{k}_1$, $\mathbf{k}_2$, and $\mathbf{n}$ lie in the same plane; let it be the vertical $xz$-plane. Then,

$$p_m^p = \frac{1}{\left|1 - \dfrac{sA}{\rho}\right|} p_1 \sin \chi. \quad (10)$$

where $p_m$ is the period of the moiré patterns (moiré period), the ratio $\rho$ is equivalently redefined as $\rho = p_2/p_1$ with $p_1$ and $p_2$ being periods of gratings.

In the case of equiangular gratings, $A = 1$, and the moiré period is

$$p_m^e = \frac{1}{\left|1 - \dfrac{s}{\rho}\right|} p_1 \sin \chi. \quad (11)$$

In the case of identical equiangular gratings, $\rho = 1$, $A = 1$ and

$$p_m^i = \frac{1}{|1 - s|} p_1 \sin \chi = \frac{l}{|\Delta l|} p_1 \sin \chi. \quad (12)$$

Another form to describe the moiré patterns is the dimensionless moiré magnification factor (magnification factor for periods) defined as follows,

$$\mu = \frac{p_m}{p_1}. \quad (13)$$

The above magnification factor is introduced similarly to [14]. In Eqs. (10) – (12) for the plane and equiangular gratings, the direction of the moiré vector is obvious, while the moiré magnification factors are

$$\mu^p = \frac{1}{\left|1 - \dfrac{sA}{\rho}\right|} \sin \chi, \quad (14)$$

$$\mu^e = \frac{1}{\left|1 - \frac{s}{\rho}\right|} \sin \chi, \quad (15)$$

$$\mu^i = \frac{l}{|\Delta l|} \sin \chi. \quad (16)$$

According to Eq. (9), the moiré wavevector is the vector sum of the wavevector of the first grating and the rescaled wavevector of the second grating. Then we follow [1] and obtain the expressions for the moiré angle and the moiré factor in the scalar form,

$$\tan \varphi_m^e = \frac{\sin \alpha}{\rho - \cos \alpha}, \quad (17)$$

$$\mu^e = \frac{\rho}{\sqrt{1 + \rho^2 - 2\rho \cos \alpha}}, \quad (18)$$

where $\alpha$ is the angle between the coplanar tangential vectors. For the identical coplanar gratings, Eqs. (17) and (18) are simplified,

$$\tan \varphi_m^e = \frac{1}{\tan \alpha/2}, \quad (19)$$

$$\mu^e = \frac{1}{2 \sin \alpha/2}. \quad (20)$$

The inverse relation between the tangents in Eq. (19) means that in this case, the moiré angle is the half angle between the vectors, but the direction of the moiré wavevector is orthogonal to the sum of the wavevectors; this corresponds to the algebraic summation with the coefficients +1 and -1 mentioned in the formal description of Eq. (1).

### B. Cylinder (paraxial approximation)

Now let's can calculate the period of the moiré patterns appearing in the screen $S$, when the rays pass a regular volumetric three-dimensional object, a single-walled cylinder, as shown in Fig. 2.

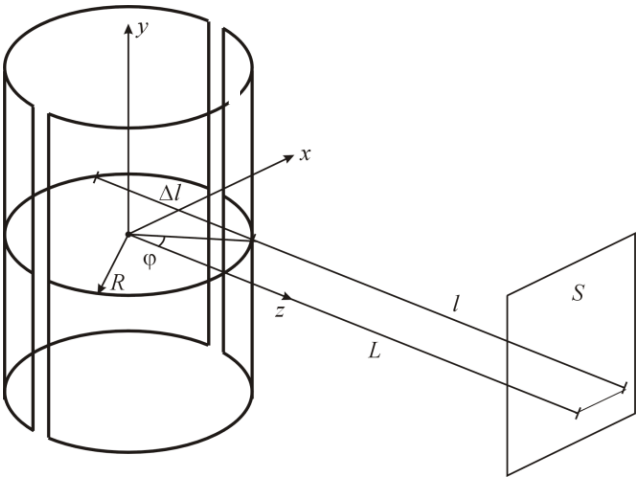

Fig. 2. Spatial object observed through both halves.

The vector LOS is perpendicular to the axis of the cylinder, $L$ is the distance from the screen to the center of the circle, and the LOS vector is directed to the origin. In the cylinder, the angles between the wavevector and LOS are equal in both halves of the cylinder, i.e., this is the orthogonal projection of the identical equiangular gratings onto the screen $S$.

The distance $\Delta l$ between the gratings along LOS is equal to the chord of the circle. Basing on the equation of the cylinder, the horizontal chord at $x$ can be written as

$$\Delta l = 2\sqrt{R^2 - x^2}. \quad (21)$$

The distance $l$ from the observer to the first grating is $L$ minus one half of the chord,

$$l = L - \frac{\Delta l}{2} = L - \sqrt{R^2 - x^2}. \quad (22)$$

Equations (21) and (22) can be rewritten in terms of the relative dimensionless variables, i.e. the ratios of the linear variables to the radius of the cylinder. Such variables will be denoted in this paper by the prime sign, for example, $L' = L/R$. With using the parametric angle $\varphi = \arcsin(x')$, one can obtain

$$\Delta l' = 2 \cos \varphi, \quad (23)$$

$$l' = L' - \cos \varphi. \quad (24)$$

To calculate the moiré factor, the orientation of the wavevector of the grating has to be taken into account. There are two types of gratings on the surface of cylinder: the vertical wavevector orthogonal to the base of the cylinder (the grid lines orthogonal to the $y$-axis) shown in Fig. 3(a) and the horizontal wavevector parallel to the base (the lines parallel to the $y$-axis) shown in Fig. 3(b). The former grating consists of the identical circles uniformly displaced along the $y$-axis; the latter consists of the vertical lines uniformly distributed along the base of the cylinder. These two cases are well-separated (this fact is confirmed experimentally in [15]) and therefore can be considered in isolation from one other.

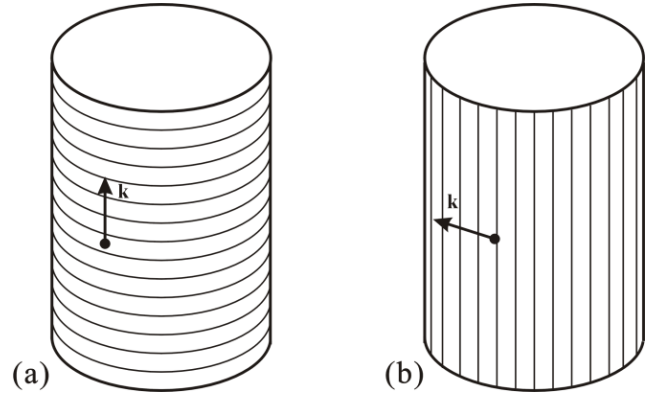

Fig. 3. Wavevectors of two types of gratings on surface of cylinder: (a) vertical wavevector; (b) horizontal wavevector (in $xz$-plane)

Both types of gratings shown in Fig. 3 are the identical equiangular gratings for which Eq. (16) is valid. Consequently, in the vertical case Fig. 3(a), $\chi = 90°$; while in the horizontal case Fig. 3(b), $\chi = 90° - \varphi$; thus $\sin \chi = \cos \varphi$. Therefore, the moiré factors of these cases are as follows,

$$\mu_v = \frac{1}{2}\left(\frac{L'}{\cos\varphi} - 1\right). \quad \textbf{(25)}$$

$$\mu_h = \frac{1}{2}(L' - \cos\varphi). \quad \textbf{(26)}$$

### C. Dual meshes

The hexagonal layout is one of the most typical arrangements in layered carbon nano-particles; it is tightly connected to the hexagonal structure of the graphene layer. Such structure can be geometrically considered as a tesselation (surface tiling).

Any regular or semiregular tesselation has a dual counterpart [16]. The vertices of one tesselation correspond to the faces of the dual tesselation and v.v. Therefore, the number of faces in one tesselation is equal to the number of faces in another; the numbers of the faces and vertices are also identical. The hexagonal tesselation and the triangular one are dual tesselations, see Fig. 4.

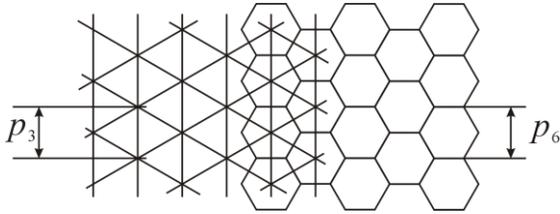

Fig. 4. Dual tesselations.

A regular tesselation considered as an image contains repeated templates. In the content of the current article, we may refer tesselations to as gratings or grids. The distance to the next repeated template depends on direction. The shortest distance is a period; in the two-dimensional case, the period is coupled with the corresponding direction. The periods of the dual gratings $p_3$ and $p_6$ are equal,

$$p_3 = p_6 = a. \quad \textbf{(27)}$$

The moiré effect is mostly caused by the interaction between the lowest spectral components. Although the spectra of the dual grids are not identical (see illustration in Fig. 5), the layout of the lowest components near the origin is the same for both dual spectra (a dashed circle in Fig. 5).

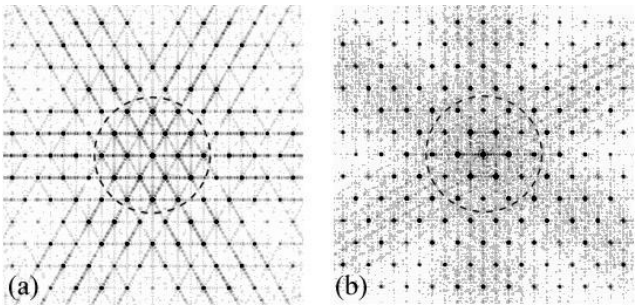

Fig. 5. Calculated two-dimensional spectra of dual tesselations. (The origin of the spectral domain is in the center of each image.)

These encircled components define the moiré effect in the most situations. Therefore, the moiré effect in the dual gratings is similar; particularly, the periods of the hexagonal moiré and of each triangular sub-structure are identical. It means that the moiré effect in the chiral SWNT can be investigated with using the triangular grid which seems to be easier for the formal considerations than the hexagonal one.

## 3. SIMULATION OF MOIRE PATTERNS IN SWNT

### A. Triangular (and hexagonal) mesh

The simulated moiré patterns are obtained by superposing the images of the gratings, as in [1], [17]. Basing on the assumption about the radius of the visibility circle and using the normalized amplitudes, the images are correspondingly filtered and re-leveled. On this way, the structure (layout, shape, and period) of the moiré patterns is not changed. Examples of simulated moiré patterns are shown in Fig. 6.

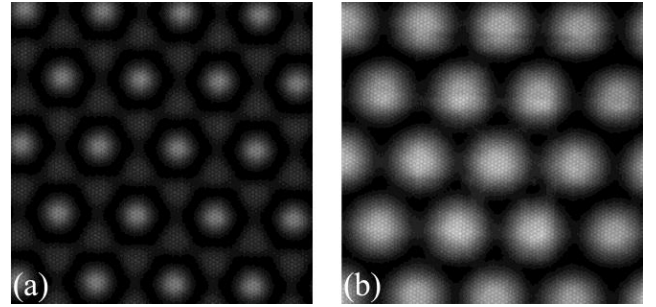

Fig. 6. Moiré effect in coplanar dual gratings satisfying Eqs. (17), (18), and (27). ($\rho = 1, A = 1$).

All simulated images in this paper do not represent the singular moiré states [1] with the in finite moiré period, rather the deviated ones. In Fig, 6, the rotational deviation is 3°; in Fig. 8 below, the size deviation is 5%.

According to [17], the highest probability of the moiré effect in square grid is reached at the rational angles with small numbers, both less than 5. Geometrically, this happens when the lines of the second grating cross the origin and the closest to the origin nodes of the first grating. These nodes define the singular moiré states [1]; the nodes can be numbered using the integer numbers $m$ and $n$. Several such nodes are shown in Fig. 7 by bold dots.

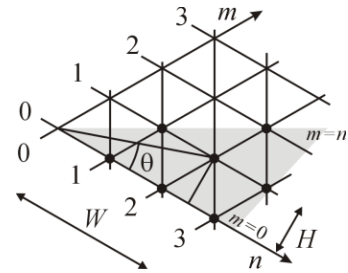

Fig. 7. Closest rational angles (bold dots) of the triangular grid.

The symmetry of the problem suggests that the moiré effect with an angle $\beta$ between the gratings is the same as at the angle $\pi/3 - \beta$, see Fig. 4. Particularly, it is sufficient to consider the angular sector $\pi/6$ between the lines $m = n$ and $m = 0$; this sector is shown in gray in Fig. 7. The nodes within this sector satisfy the condition $m \leq n$. The boundary lines correspond to the armchair and zigzag configurations of the hexagonal grid. The nodes of the dual hexagonal grid are located at the same points (if the common displacement vector is neglected).

The triangle built on the $n$-axis and the $(m, n)$-th node of the grid has dimensions

$$H = m\frac{\sqrt{3}}{2}b, \quad W = \left(n + \frac{m}{2}\right)b, \quad \textbf{(28)}$$

where $H$ and $W$ are the height and the width of the triangle, $b$ is the side of the basic triangle. Note that the period of the triangular grid $p_3 = H(1) = b \cdot 3^{1/2}/2$. Then, the angle from the $n$-axis and the relative length of the hypotenuse (actually it is the ratio of the periods of the gratings, i.e., the parameter $\rho$ defined by Eq. (7)) are as follows

$$\tan\theta = \frac{H}{W} = \frac{\sqrt{3}m}{m + 2n}, \quad \textbf{(29)}$$

$$\rho = \frac{\sqrt{H^2 + W^2}}{p_3} = \frac{2}{\sqrt{3}}\sqrt{m^2 + mn + n^2}. \quad \textbf{(30)}$$

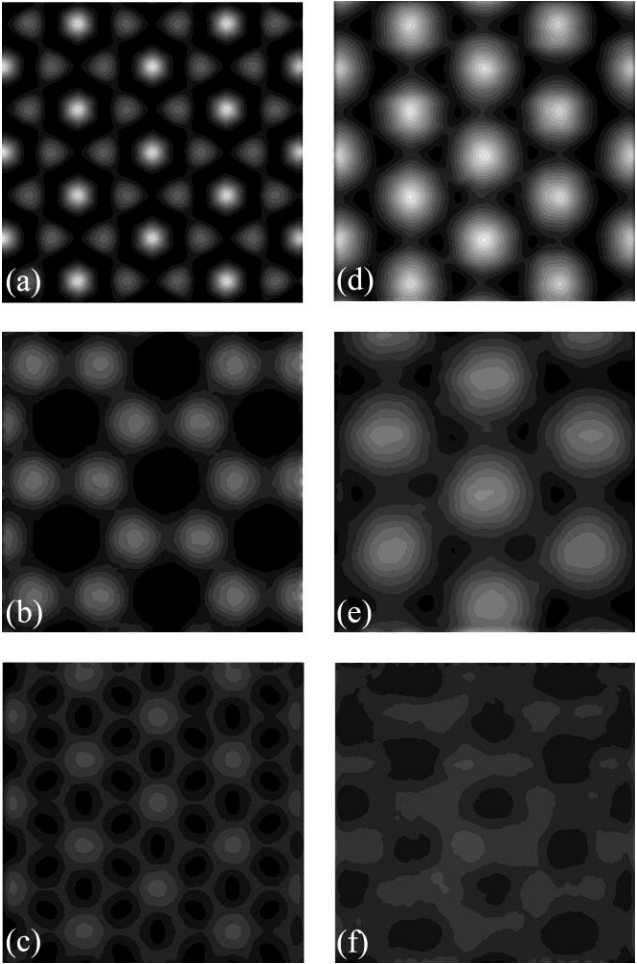

Fig. 8. Simulation of three closest moirés in dual (triangle on the left, hexagonal on the right) grids $(m, n)$ = (0,1), (1,1), and (0,2). The deviation from theoretical values $\rho$ = 1, 1.73, and 2 is 5%; the angles are 0, 30°, and 0.

Among four nodes of the triangular grid closest to the origin, there are two nodes on the $n$-axis with $n = 1$ and $n = 2$. For them,

$$\rho_{01} = 1, \quad \rho_{02} = 2. \quad \textbf{(31)}$$

Two other non-zero angles correspond to $(m, n)$ = (1, 1) and (1, 2). In these cases,

$$\rho_{11} = \sqrt{3} = 1.73, \quad \rho_{12} = \sqrt{7} = 2.65. \quad \textbf{(32)}$$

The corresponding angles 30° and 19.1° are obtained from Eq. (29) (the tangents are $3^{1/2}/3$ and $3^{1/2}/5$). The former layout partially coincides with the slanted lines of the triangular grid.

Figure 8 shows the simulated moiré patterns in dual grids. It can be seen that in the dual grids, the moiré patterns precisely well correspond to each other, at least in the layout of the patterns and in their period. However, it should be noted that there are three overlapped triangular sub-structures in the triangular case, see Fig. 8(a). Figure 8 also illustrates that the higher spectral components (with higher indices $m$ and $n$) play less role; the closest and visually strongest moiré patterns are observed when $(m, n)$ = (0,1) and (1,1).

In simulation, the modeled gratings have the rectangular profile. However, the profile of the graphene layer [18] is rather smooth, i.e., it contains fewer spectral components. Therefore for the moiré effect in the plane graphene gratings as well as in SWNT, only the lowest spectral components are essential. Practically, we assume that it is enough to consider the first spectral harmonic.

**B. Slanted (wrapped) cylinder**

The chiral nano-particles can be thought as a hexagonal mesh bent into a cylinder (wrapped around it) comprising a single-walled nanotube (SWNT). The seamless condition makes certain restrictions for the mesh to be bent, and a variety of mesh layouts can be described by two integer indices $m$ and $n$. For the triangular grid, the chiral indices were actually introduced in Sec. 3A.

The vertical components of the wavevectors of gratings on the frontal half of the cylinder and on its rear half are opposite, i.e., the slant angle on the rear surface is the negative angle $\theta$ on the frontal surface. Therefore, the angle between the wavevectors near the axis of the cylinder is $2\theta$, as schematically shown in Fig. 9.

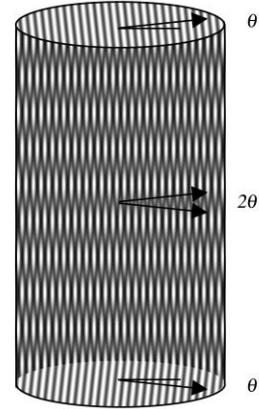

Fig. 9. Scheme of angles at the front and rear surfaces of cylinder.

However, for the moiré effect we do not add the wavevectors directly, rather we use their algebraic sum with coefficients -1 and +1. Therefore for the slanted identical gratings, the moiré vector is directed down, i.e., it is orthogonal to the sum of the wavevectors of the gratings. (Although, the moiré wavevector is not shown in Fig. 9, its direction is clear from the moiré patterns themselves.)

The diameter and the chiral angle of the SWNT are known from many papers and textbooks, e.g., [19]. Namely, the diameter can be calculated basing on the chiral indices

$$d = \frac{a}{\pi}\sqrt{n^2 + nm + m^2}, \quad (33)$$

where $0 \leq m \leq n$, and $a$ is the period of the hexagonal tesselation defined in Eq. (27). Note that $a = \sigma \cdot 3^{1/2}$, where $\sigma$ is the side of the hexagon (the bond length). There are two fundamental configurations of bent hexagonal meshes on the cylinder, the zigzag and the armchair configurations with $m = 0$ and $m = n$, resp. Other configurations have intermediate indices.

For the identical coplanar gratings, the direction of the moiré wavevector is equal to the half angle between the wavevectors in the linear combination with coefficients +1 and -1. In the case shown in the scheme Fig. 9, the on-axis moiré factor can be derived from Eq. (20),

$$\mu = \frac{1}{2\sin 2\theta/2} = \frac{1}{2\sin\theta}. \quad (34)$$

The sinus of the chiral angle $\theta$ (measured from the zigzag direction) can be written (as in [19])

$$\sin\theta = \frac{\sqrt{3}m}{2\sqrt{m^2+mn+n^2}}. \quad (35)$$

Consequently, the moiré period depends on the indices as follows,

$$T_m = \frac{a}{\sqrt{3}m}\sqrt{m^2+mn+n^2}. \quad (36)$$

Thus, the moiré patterns in the chiral SWNT have discrete periods and discrete angles. In particular, when $m = n$, the period $T_m = a$; while when $m = 0$, $T_m = \infty$. In the intermediate configurations, the moiré period varies between these values. Therefore, the moiré patterns near the axis of the cylinder can be modeled by two coplanar plane layers at proper angles. Examples of the simulated patterns are shown in Fig. 10.

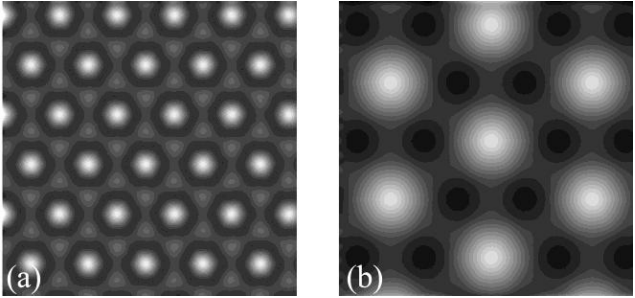

Fig. 10. The simulated moiré patterns in plane dual gratings installed at the angle 28°.

The chiral angle 28° in Fig. 10 corresponds to the SWNT with $(m, n) = (9, 8)$. Besides, the images in Fig. 6 (if rotated counter-clockwise by 1.5°) correspond to the moiré patterns in CWNT with $(m, n) = (12, 11)$.

In contrast to the infinite plane layers, the observation the moiré patterns in cylinders with the period larger than the finite radius of the cylinder is practically impossible. This yields the following condition,

$$m > \frac{2\pi}{\sqrt{3}} = 3.63, \quad (37)$$

or (because $m$ is integer),

$$m \geq 4. \quad (38)$$

Equation (38) represents the condition to observe the single moiré fringe. However, a single fringe is practically not enough to measure the period accurately. Then, $N$ fringes per radius (the central area of cylinder) can be observed when

$$m \geq 4N. \quad (39)$$

Hence, several (two or more) moiré fringes can only be observed with $m$ larger than approx. ten.

Among other things, one may find basing on the symmetry that except for the moiré patterns with their origin at $\theta = 0$, there are also the symmetrical moiré patterns with the origin at the chiral angle $\theta = \pi/6$. For them, the moiré factor is

$$\mu = \frac{1}{2\sin(\pi/6 - \theta)}. \quad (40)$$

Two symmetric patterns in the cylinder are graphically illustrated in Fig. 11, where the theoretical curves for the moiré period by Eqs. (34) and (40) are shown together with the experimental measurements. The normalized root-mean-square difference between the theoretical end experimental data shown in Fig. 11 is about 5%.

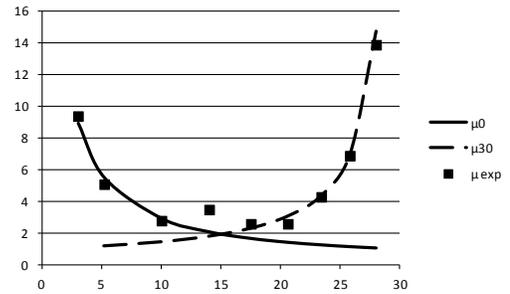

Fig. 11. The theoretical and experimental moiré period vs. chiral angle for two branches in the hexagonal grid.

The period of the symmetric pattern can be also expressed in terms of the chiral indices $m$ and $n$,

$$T_{mS} = \frac{a}{n-m}\sqrt{m^2+mn+n^2}. \quad (41)$$

The pattern Eq. (36) has the longer period below 10°, the pattern Eq. (41) prevails above 20°. The orientation of the symmetric moiré patterns is different: the horizontal hexagon in the case of Eq. (36), see Fig. 12(a) and the vertical hexagon in the case of Eq. (41), see Fig. 12(b).

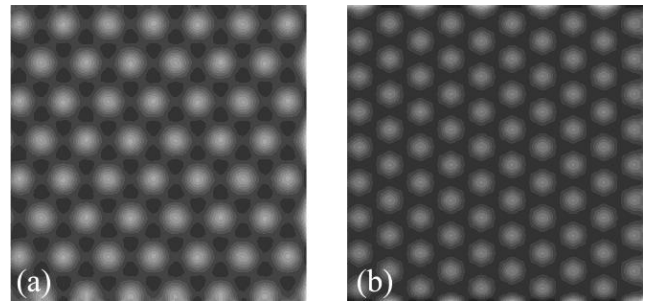

Fig. 12. The simulated symmetrical moiré patterns (hexagonal grid, chiral angles 5° and 23°).

The period of the observed moiré patterns is the maximum of two Eqs. (36) and (41); it has the minimum when

$$m_{\min} = \frac{\sqrt{3}-1}{2}n = 0.37n. \quad \textbf{(42)}$$

The graphs Fig. 13 show the diameter of the nanotube and the period of the moiré patterns as a functions of $m$.

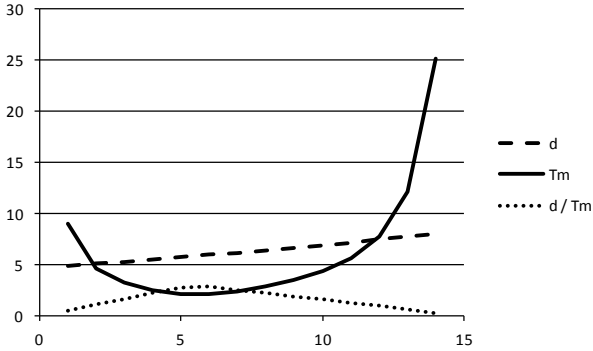

Fig. 13. Diameter of nanotube $d$, moiré period $T_m$ and number of fringes per diameter as functions of $m$ (sizes normalized by $a$; $n$ = 15).

The graphs in Fig. 13 were calculated from Eqs. (33), (36), and (41). The graphs show that at the edges of the angular range, the moiré period is typically larger than the diameter, and thus, the observation of the moiré patterns is practically impossible. However, when $m$ is close to $0.37n$, the moiré period can be smaller than the diameter. This makes the intermediate configurations ($0 < m < n$) preferable for the observation of the moiré patterns in nanotubes.

Note that among $N$ moiré periods per radius, measurable are $N$-2 periods only. This is because near the edges of the cylinder, the period of the moiré patterns cannot be measured exactly due to the near–edge distortion of the period by Eq. (26) and to the non-zero thickness of the physical mesh. Therefore, a few moiré periods, say, one or two periods per diameter, are practically non-measurable. The maximum $x$-value for some reliable measurements is as follows,

$$x'_{\max} = \frac{N-2}{N}. \quad \textbf{(43)}$$

For example, $x'$ = 0.6 for $N$ = 5. The condition Eq. (43) practically coincides with the above calculation of the number of fringes per radius; because $2x_{max}$ = 1.2 is close to 1, especially for Eqs. (39) and (43), which are essentially estimations. For illustration, refer to Fig. 14. It is also convenient in cylinders that within the range ($-x_{max}$, $+x_{max}$), the period does not deviate essentially from the on-axis value.

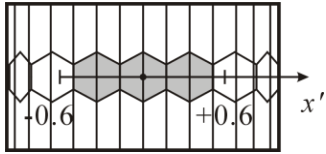

Fig. 14. Measurable moiré periods in cylinder (grayed hexagons).

In experiments we used the printed binary gratings (pure black and pure white colors only) with the relative width of the lines about 50%. According to [20], such width provides the maximum amplitude of the moiré patterns. The experimental illustrations of the moiré effect in the dual hexagonal and triangular grids are shown in Fig. 15.

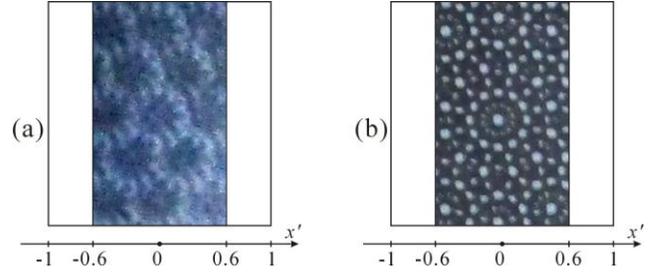

Fig. 15. Experimental photographs of the near-axis ($|x'| < 0.6$) moiré patterns with $n \geq m > 10$ (obtained with using the printed plane triangular and hexagonal grids); $(m, n)$ = (15.6), chiral angle 16.1°.

To summarize, the moiré patterns can be observed in chiral nanotubes with $m$ larger than ten, and in this case, there are several periods of patterns across the diameter. The picture of the moiré patterns displays the initial (non-rotated) grid for any chiral angle.

## 4. DISCUSSION AND CONCLUSIONS

The hexagonal grid consists of identically located identical hexagons only, and the moiré patterns reproduce the hexagonal grid. However, the triangular grid consists of two sorts of triangles, oriented to the left and to the right, see Fig. 4. Because of that, details of the moiré patterns in the triangular grid can be classified into the left- or right-oriented triangles observed through the opposite triangles and of the either triangles observed through triangles of the same type. These combinations of triangles comprise four overlapped triangular sub-structures (two of which coincide), see Fig. 8(a). The moiré period is the same in all four cases.

As a result, the triangular moiré consists of 3 overlapped triangular sub-structures which may appear similarly to each other. The three sub-structures are recognizable in Figs. 6(a) and 8(a). If the difference between them is not taken into account, and they are occasionally treated as identical, then the seeming period is $\sqrt{3}$ times shorter.

It is interesting that the moiré factor for the orthogonal equiangular gratings ($\chi$ = 90°) is

$$\mu^{eo} = \frac{1}{\left|1 - \dfrac{s}{\rho}\right|}. \quad \textbf{(44)}$$

On the other hand, the well-known magnification factor of the thin lens [21] can be written in the following form

$$M = \frac{d_i}{d_o} = -\frac{1}{1 - \dfrac{d}{f}}, \quad \textbf{(45)}$$

where $d_i$ and $d_o$ are the distances to the image and to the object; $d$ is the distance to the object, and $f$ is the focal distance of the lens, while the sign of the expression tells the difference between the real lens images and virtual ones.

Therefore, the ratio $\rho$ in Eq. (44) could be an equivalent of the focal distance of the "moiré lens", while $s$ might correspond to the distance between the "lens" and the object. Consequently, the moiré interaction between the orthogonal equiangular gratings could formally represent a lens. The ratio of periods defines the focal distance of such "moiré lens", while the ratio of distances to the first and to the second gratings defines the effective distance to the object. In other words, the inner structure (periods) of layers defines the focal length, while their layout (the distances) defines the effective distance.

By the re-projection of the second grating, we imply that the visual image of the moiré patterns lies in the tangential plane of the first grating. However strictly speaking, this might not be a very general case, see, e.g., [22], where the visual image is perceived between two physical planes (depending on brightness). It is also important that that projected quantity is the length, but not the wavevector; then the wavevector is calculated basing on the projected period.

When considering the moiré effect along the vertical axis of the cylinder, the moiré factors obtained for the horizontal plane should be multiplied by the sine of the angle between the LOS and the vertical axis, similarly to Eq. (15) for the equiangular surfaces.

It is proven in this paper that the triangular simulation of the moiré effect is quite similar to the hexagonal one (the hexagonal grid is mostly provided for comparison). Basically, the difference is only in the shape of the moiré cells, while the periods and orientations of the patterns are the same. The normalized root-mean-square difference between the theoretical end experimental data is about 5%.

A further research of the moiré effect in the chiral SWNT can be made with using the triangular grid only, which is easier in implementation and in consideration, because it is just linear combinations of solid line gratings; this is not the case for the hexagonal grid. In the further research, an advanced theory based on the central projection should be built, correspondingly re-arranged experiment would be needed. An additional research is needed to locate exactly the visual image of the moiré patterns between the layers.

In the current paper we built a theory for the moiré effect in the single-layer particles (chiral SWNT) basing on the orthogonal projection; several particular cases are considered; we propose using of the dual triangular grid for the simulation and the analytical development of the moiré patterns in the hexagonal SWNT; the formula for the moiré period versus the chiral angle is obtained; we determined the chiral indices and the configurations of the mostly expected moiré patterns. Also, we proposed the moiré magnification factor and found its similarity to the lens magnification in the case of the orthogonal equiangular gratings. The results of this research can be applied to the meshed cylinders in general, as well as to the chiral nanoparticles in particular.